\documentclass[%
 reprint,
superscriptaddress,
 amsmath,amssymb,
 aps,
]{revtex4-2}

\usepackage{graphicx}
\usepackage{subfigure} 
\usepackage{dcolumn}
\usepackage{bm}
\usepackage[colorlinks=true,
            linkcolor=blue,
            citecolor=blue,
            urlcolor=blue]{hyperref}
\usepackage{xcolor}

\begin{document}

\preprint{APS/123-QED}

\title{Eigenstate Transitions, Duality, and Anomalous Diffusion in a Quasiperiodic Qi-Wu-Zhang Chern Insulator}

\author{Tengming Lou}
\affiliation{Center for Advanced Quantum Studies, School of Physics and Astronomy, Beijing Normal University, Beijing 100875, China}
\author{Haiyang Wang}
\affiliation{Center for Advanced Quantum Studies, School of Physics and Astronomy, Beijing Normal University, Beijing 100875, China}
\author{Haijiao Ji}
\affiliation{Tsung-Dao Lee Institute, Shanghai Jiao Tong University, Shanghai 201210, China}
\author{Haiwen Liu}
\email{haiwen.liu@bnu.edu.cn}
\affiliation{Center for Advanced Quantum Studies, School of Physics and Astronomy, Beijing Normal University, Beijing 100875, China}
\affiliation{Key Laboratory of Multiscale Spin Physics, Ministry of Education, Beijing 100875, China}
\affiliation{Interdisciplinary Center for Theoretical Physics and Information Sciences, Fudan University, Shanghai 200433, China}


\begin{abstract}
Quasiperiodic systems usually interpolate between extended, critical, and localized states as the quasiperiodic modulation is increased. Here we show that the magnetic Qi-Wu-Zhang Chern-insulator model realizes a distinct full-spectrum transition in which localization is avoided. For an irrational magnetic flux, the two-dimensional model reduces to a spinor quasiperiodic chain with a matrix onsite modulation controlled by the hopping amplitude \(t_x\). When \(|m+2|>t_y\), increasing \(t_x\) produces the conventional extended-critical-localized sequence with a critical line at \(t_x=t_y\). In contrast, when \(|m+2|\leqslant t_y\), the system changes from an extended phase to a critical phase at \(t_x=|m+2|\) and remains critical even for stronger quasiperiodic modulation. Finite-size scaling of the average inverse participation ratio gives \(\overline{\mathrm{IPR}}\sim q^{-\alpha}\) with \(0<\alpha<1\) throughout this persistent critical regime. A dual transformation exchanging \(t_x\) and \(t_y\), together with a Lyapunov-exponent analysis, explains the phase diagram. Wave-packet dynamics further distinguish ballistic, anomalous-diffusive, and localized regimes. These results identify magnetic Chern-insulator systems as a natural platform for robust criticality and anomalous quantum transport.
\end{abstract}

\maketitle


\section{Introduction}\label{sec1}
The Hofstadter problem shows that a lattice electron in a perpendicular magnetic field can acquire a fractal spectrum when the magnetic flux per unit cell is incommensurate with the lattice period~\cite{PhysRevB.14.2239,Wannier1978,P_Streda_1982}. After a partial Fourier transformation, the magnetic phase appears as a quasiperiodic modulation, connecting the two-dimensional Hofstadter setting to the Aubry-Andr\'{e}-Harper class of one-dimensional quasiperiodic models~\cite{P_G_Harper_1955,aubry1980analyticity,Peierls1933763}. These systems provide a controlled setting in which extended, critical, and localized eigenstates can be studied without uncorrelated random disorder.

In one-dimensional quasiperiodic systems, localization physics is richer than in both periodic crystals and uncorrelated disordered systems. The Aubry-Andr\'{e} model has a self-dual localization transition, and more general spinful or multicomponent quasiperiodic chains can host critical phases, mobility edges, and coexistence regimes~\cite{aubry1980analyticity,PhysRevLett.50.1870,ZHOU20261654,PhysRevLett.125.196604,PhysRevLett.131.176401}. Critical states are particularly important because they are neither Bloch extended nor exponentially localized; instead, their wave functions show multifractal scaling and their dynamics are generally anomalous~\cite{PhysRevLett.50.1873,PhysRevB.29.1394,10.1098/rspa.1984.0016,PhysRevA.33.1141,PhysRevA.37.1345,PhysRevB.38.5811,PhysRevB.34.2041,HiramotoKohmoto1992,PhysRevLett.131.186303,Goblot2020832}.

Here we ask whether a natural two-dimensional Chern-insulator model in an irrational magnetic field can realize localization physics that is not simply inherited from scalar Aubry-Andr\'{e} chains. We focus on the Qi-Wu-Zhang (QWZ) Chern insulator~\cite{PhysRevB.74.085308}. In a Landau gauge, the magnetic QWZ model reduces to a one-dimensional spinor quasiperiodic chain whose onsite potential is a matrix rather than a scalar. The parameters \(t_x\), \(t_y\), and \(m+2\) respectively control the quasiperiodic modulation, the intercell hopping along the reduced chain, and the topological mass. This structure gives an experimentally and conceptually simple route to quasiperiodic criticality in a Chern-insulator setting.

\begin{figure*}[t]
\centering
\includegraphics[width=0.9\linewidth]{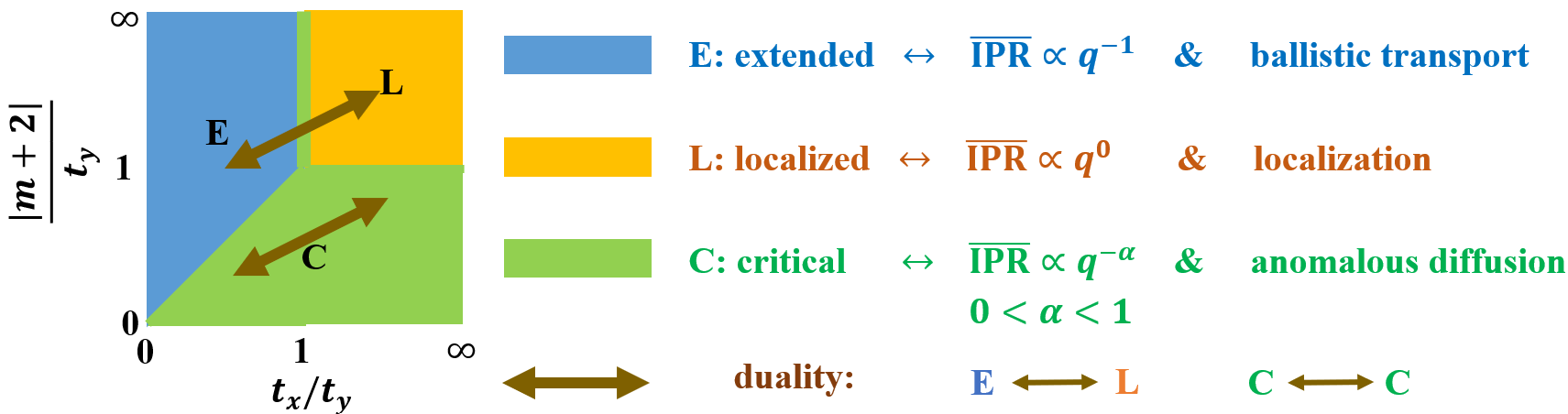}
\caption{Phase diagram and diagnostics of the magnetic QWZ model in the reduced quasiperiodic representation. Extended, localized, and critical regimes are denoted by E, L, and C. The conventional regime \(|m+2|>t_y\) contains the extended-critical-localized sequence across \(t_x=t_y\), whereas the regime \(|m+2|\leqslant t_y\) contains a persistent critical phase for \(t_x\geqslant |m+2|\). The average inverse participation ratio scales as \(\overline{\mathrm{IPR}}\propto q^{-1}\), \(q^{0}\), and \(q^{-\alpha}\) with \(0<\alpha<1\) in the extended, localized, and critical regimes, respectively. Wave-packet dynamics distinguish the same regimes through ballistic spreading, localization, and anomalous diffusion.}
\label{fig1}
\end{figure*}

Our main result is a persistent critical phase, as shown in Fig.~\ref{fig1}. When \(|m+2|>t_y\), increasing \(t_x\) yields the conventional extended-critical-localized sequence, with the critical line at \(t_x=t_y\). In contrast, when \(|m+2|\leqslant t_y\), the system undergoes an extended-to-critical transition at \(t_x=|m+2|\) and then remains critical even for stronger quasiperiodic modulation. Thus, in this regime, stronger modulation does not produce exponential localization. We characterize the transition by finite-size scaling of the average inverse participation ratio, explain the phase boundaries using a duality that exchanges \(t_x\) and \(t_y\), and confirm the transport consequences through wave-packet dynamics.

This result is also motivated by recent work showing that Chern-band topology can enrich Hofstadter spectra and produce anomalous open-orbital subbands beyond the conventional Diophantine window~\cite{Ji2025Anomalous,PhysRevLett.49.405}. The present work addresses the complementary eigenstate and dynamical problem: how the magnetic QWZ model localizes, remains critical, or transports a wave packet as the quasiperiodic modulation is tuned. The paper is organized as follows. Sec.~\ref{sec2} derives the reduced quasiperiodic model. Sec.~\ref{sec3} presents the eigenstate transitions and inverse-participation-ratio scaling. Sec.~\ref{sec4} develops the duality and Lyapunov-exponent interpretation. Sec.~\ref{sec5} studies wave-packet dynamics, and Sec.~\ref{sec6} summarizes the results and open questions.

\section{\label{sec2}QWZ model under a magnetic field}
In this section, we present the treatment of the QWZ model under a magnetic field. 
Considering the nearest-neighbor hopping, the tight-binding Hamiltonian of the QWZ model under a perpendicular magnetic field \(B\hat{z}\) is given by~\cite{PhysRevB.14.2239,PhysRevB.74.085308,Ji2025Anomalous}
\begin{widetext}
\begin{equation}
H = \sum_{n_x,n_y} c_{n_x,n_y}^\dagger M c_{n_x,n_y} - \sum_{n_x,n_y} \left( c_{n_x,n_y}^\dagger T_x c_{n_x+1,n_y} + c_{n_x,n_y}^\dagger T_y c_{n_x,n_y+1} + \text{h.c.} \right),
\end{equation}
\end{widetext}
where \(c_{n_x,n_y}^\dagger\) and \(c_{n_x,n_y}\) represent the creation and annihilation operators with two components, respectively; 
\(M = (m+2)\sigma_z\) denotes the onsite mass term with topological 
mass \(m\); \(T_{x(y)} = \dfrac{t_{x(y)}}{2}(\sigma_z + i\sigma_{x(y)}) e^{i\varphi_{x(y)}}\) is 
the nearest-neighbor hopping term with hopping amplitude \(t_{x(y)}\) and 
the Pauli matrices \(\sigma_{x,y,z}\); and \(\varphi_x\) (\(\varphi_y\)) denotes the 
Peierls phase along the \(x\)- or \(y\)-direction. 
From position \((x,y)\) to \((x',y')\), the Peierls phase is given by~\cite{Peierls1933763}
\begin{equation}
\varphi = \frac{e}{\hbar} \int_{(x,y)}^{(x',y')} \bm{A} \cdot \mathrm{d}\bm{r}.
\end{equation}

In the Landau gauge \(\bm{A} = (-By, 0)\), we obtain \(\varphi_y=0\) 
and \(\varphi_x = -2\pi \beta n_y\), where \(\beta=\dfrac{\sqrt{5}-1}{2}\) is the number of magnetic-flux quanta per unit cell. 
Due to the magnetic translation symmetry along the \(x\)-direction, 
the problem can be reduced to a 1D tight-binding model through a Fourier transform along 
the \(x\)-direction~\cite{PhysRevB.14.2239,P_G_Harper_1955,aubry1980analyticity,Ji2025Anomalous}. 
In numerical calculations, we approximate the irrational flux \(\beta\) by a rational approximant \(p/q = F_l/F_{l+1}\), where \(F_l\) denotes the \(l\)th term of the Fibonacci sequence. 
Consequently, the eigen-equation for the QWZ model under a magnetic field is given by
\begin{equation}\label{eigeneq}
E \psi(n_y) = U(n_y) \psi(n_y) - T_y \psi(n_y+1) - T_y^\dagger \psi(n_y-1),
\end{equation}
where
\(\psi(n_y)=\begin{pmatrix} \psi_A(n_y)  \\ \psi_B(n_y)  \end{pmatrix}\), 
with \(A\) and \(B\) denoting the two atomic orbitals. Here,
\begin{widetext}
\begin{equation}
U(n_y) = \begin{pmatrix} 
m+2 - t_x \cos(k_x + 2\pi \beta n_y) & t_x \sin(k_x + 2\pi \beta n_y) \\ 
t_x \sin(k_x + 2\pi \beta n_y) & -[m+2 - t_x \cos(k_x + 2\pi \beta n_y)]
\end{pmatrix},
\end{equation}
\end{widetext}
and the hopping matrix \(T_y\) is defined as
\(T_y = \dfrac{t_y}{2} \begin{pmatrix} 1 & 1 \\ -1 & -1 \end{pmatrix}\). 
After the partial Fourier transform along the \(x\)-direction, the magnetic field is encoded in the onsite matrix \(U(n_y)\). The hopping amplitude \(t_x\) sets the strength of this quasiperiodic matrix modulation, whereas \(t_y\) is the nearest-neighbor hopping along the reduced chain. The mass \(m+2\) controls the relative offset of the two orbital components. The localization problem is therefore governed by the competition among \(t_x\), \(t_y\), and \(|m+2|\), rather than by a single scalar quasiperiodic potential.

According to the magnetic Bloch condition 
\(\psi(n_y+q) = e^{ik_y q} \psi(n_y)\), 
the corresponding Hamiltonian matrix can thus be written as
\begin{equation}\label{ham_matrix}
h = \begin{pmatrix}
U(1) & -T_y & \mathbf{0} & \cdots & -T_y^\dagger e^{-ik_y q} \\
-T_y^\dagger & U(2) & -T_y & \cdots & \mathbf{0} \\
\vdots & \vdots & \ddots & \vdots & \vdots \\
\mathbf{0} & \cdots & -T_y^\dagger & U(q-1) & -T_y \\
-T_y e^{ik_y q} & \cdots & \mathbf{0} & -T_y^\dagger & U(q)
\end{pmatrix} .
\end{equation}
The quasiperiodic case is recovered in the limit \(q \to \infty\).

\section{\label{sec3}Eigenstate Transitions}
In Sec.~\ref{sec2}, we derived the Hamiltonian matrix for the QWZ model under a magnetic field. 
The eigenenergies and eigenstates can be obtained by diagonalizing the Hamiltonian in Eq.~\eqref{ham_matrix}. 

Before discussing the general case, we note that in the special limit \(t_x=0\) and \(m=-2\), the onsite potential vanishes and the system reduces to a translationally invariant chain. This limit, which corresponds to the absence of magnetic-field effects and yields extended Bloch waves with \(\overline{\mathrm{IPR}}=1/q\), is discussed in Appendix~\ref{app:no_field}.

\subsection{Two types of eigenstate transitions}
We first classify the eigenstates as functions of \(t_x/t_y\) and \((m+2)/t_y\). Throughout the numerical results below we set \(t_y=1\) and take \(t_x>0\); the general phase boundaries are restored by replacing \(m+2\) with \(|m+2|\).

For \(|m+2|>t_y\), the system exhibits the conventional quasiperiodic localization sequence. The states are extended for \(t_x<t_y\), critical on the self-dual line \(t_x=t_y\), and localized for \(t_x>t_y\). For \(|m+2|\leqslant t_y\), however, the transition is qualitatively different: the states are extended for \(t_x<|m+2|\), become critical at \(t_x=|m+2|\), and remain critical for \(t_x>|m+2|\). This second regime is the persistent critical phase emphasized in this work.

\begin{figure*}[t]
\centering
\subfigure{
  \begin{minipage}[b]{0.183\linewidth}\label{fig2a}
    \centering
    \includegraphics[width=1\linewidth]{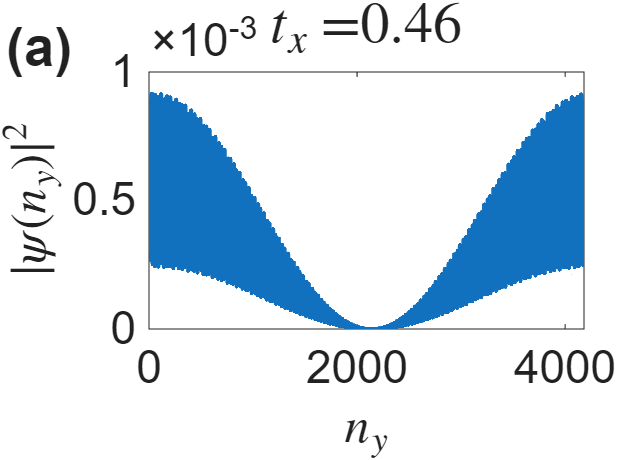}
  \end{minipage}
}
\subfigure{
  \begin{minipage}[b]{0.183\linewidth}\label{fig2b}
    \centering
    \includegraphics[width=1\linewidth]{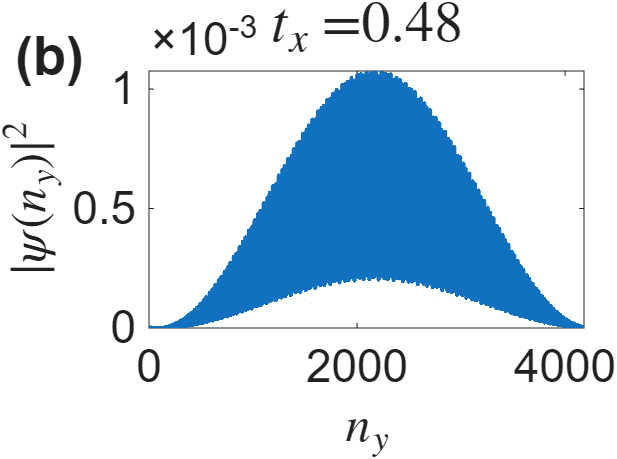}
  \end{minipage}
}
\subfigure{
  \begin{minipage}[b]{0.172\linewidth}\label{fig2c}
    \centering
    \includegraphics[width=1\linewidth]{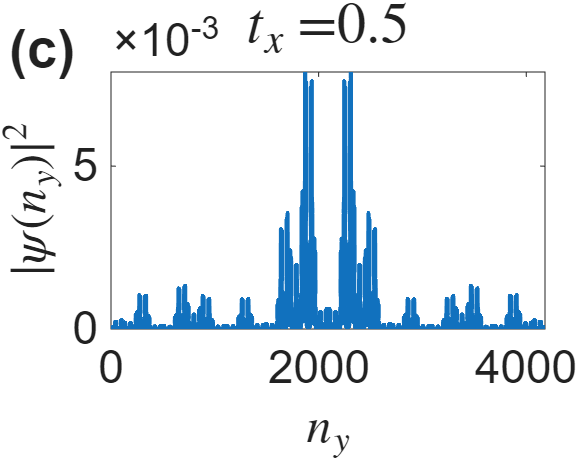}
  \end{minipage}
}
\subfigure{
  \begin{minipage}[b]{0.189\linewidth}\label{fig2d}
    \centering
    \includegraphics[width=1\linewidth]{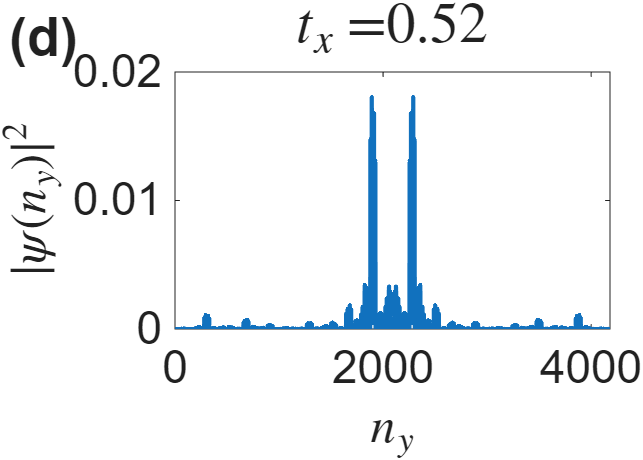}
  \end{minipage}
}
\subfigure{
  \begin{minipage}[b]{0.189\linewidth}\label{fig2e}
    \centering
    \includegraphics[width=1\linewidth]{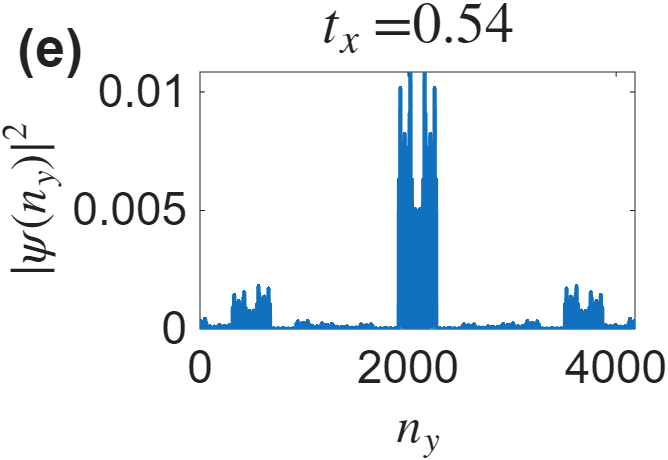}
  \end{minipage}
}
\subfigure{
  \begin{minipage}[b]{0.176\linewidth}\label{fig2f}
    \centering
    \includegraphics[width=1\linewidth]{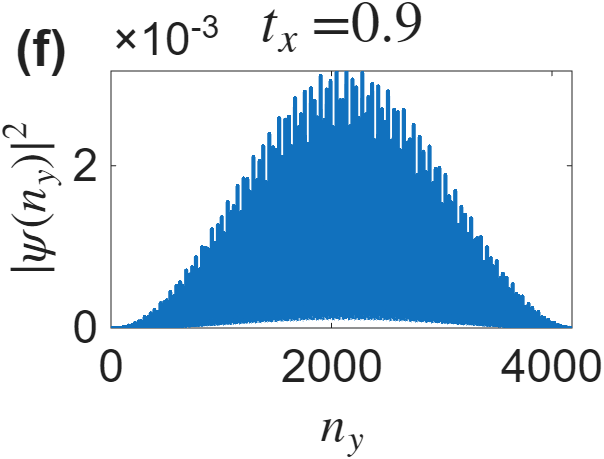}
  \end{minipage}
}
\subfigure{
  \begin{minipage}[b]{0.176\linewidth}\label{fig2g}
    \centering
    \includegraphics[width=1\linewidth]{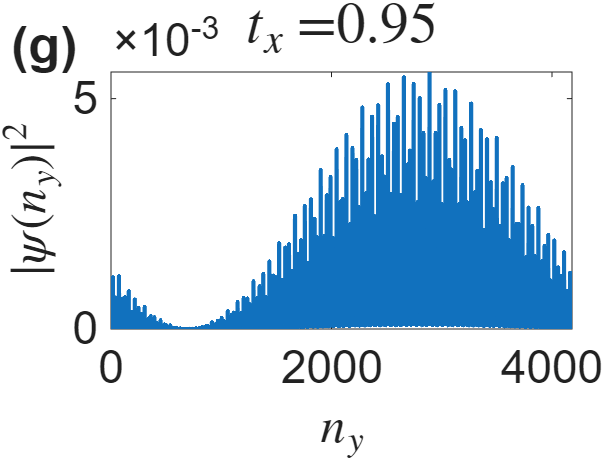}
  \end{minipage}
}
\subfigure{
  \begin{minipage}[b]{0.193\linewidth}\label{fig2h}
    \centering
    \includegraphics[width=1\linewidth]{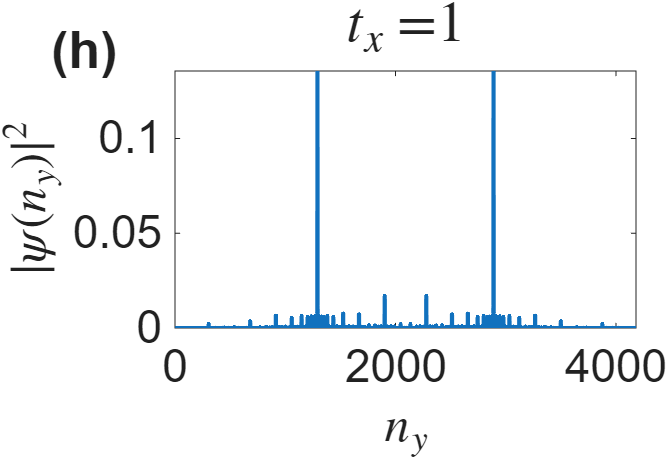}
  \end{minipage}
}
\subfigure{
  \begin{minipage}[b]{0.186\linewidth}\label{fig2i}
    \centering
    \includegraphics[width=1\linewidth]{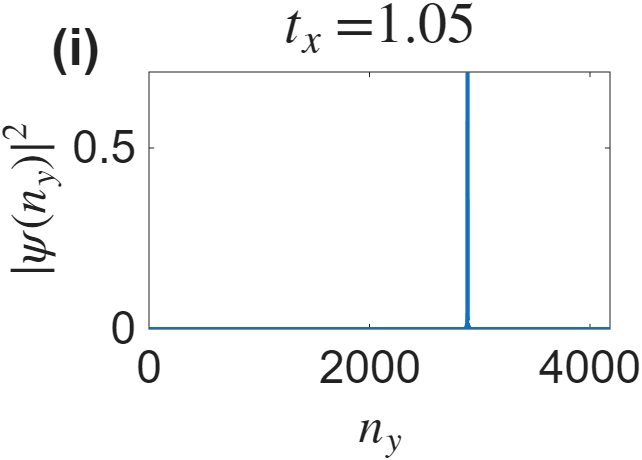}
  \end{minipage}
}
\subfigure{
  \begin{minipage}[b]{0.185\linewidth}\label{fig2j}
    \centering
    \includegraphics[width=1\linewidth]{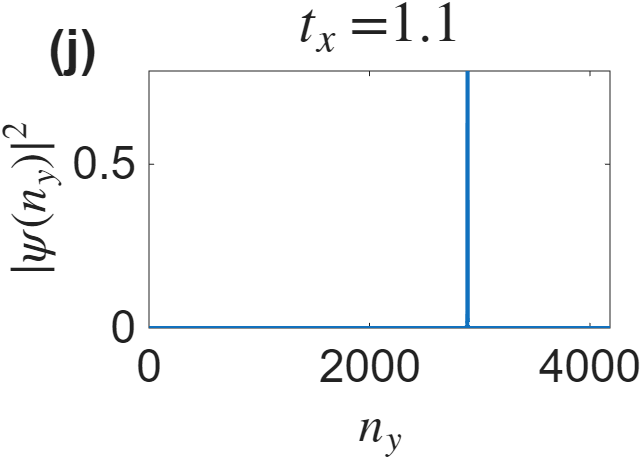}
  \end{minipage}
}
\caption{Representative probability densities \(|\psi(n_y)|^2=|\psi_A(n_y)|^2+|\psi_B(n_y)|^2\) for the magnetic QWZ model with \(q=4181\) and \(t_y=1\). Panels (a)--(e) show \(m=-1.5\): the state is extended for \(t_x<0.5\), becomes critical at \(t_x=0.5\), and remains critical for larger \(t_x\). Panels (f)--(j) show \(m=-0.5\): the state is extended for \(t_x<1\), critical at \(t_x=1\), and localized for \(t_x>1\). The critical states display multifractal spatial structure distinct from both the extended and localized limits.}
\label{fig:wf}
\end{figure*}

Representative wave functions are shown in Fig.~\ref{fig:wf}. For \(m=-1.5\), the transition occurs at \(t_x=m+2=0.5\), and the post-transition wave functions remain spatially sparse and multifractal rather than exponentially localized. For \(m=-0.5\), the system follows the conventional sequence with a critical point at \(t_x=1\) and localized states for larger \(t_x\). The visual distinction in Fig.~\ref{fig:wf} is quantified below by finite-size scaling of the average inverse participation ratio.

\subsection{Finite-size scaling of \texorpdfstring{\(\overline{\mathrm{IPR}}\)}{IPR}}
To quantify the eigenstate character, we compute the average inverse participation ratio (IPR). For the \(j\)-th normalized two-component eigenstate, the IPR is defined as~\cite{Wegner1980209}
\begin{equation}\label{IPR}
\mathrm{IPR}_{j}=\sum_{n_y=1}^{q}\left(|\psi_A^{(j)}(n_y)|^2+|\psi_B^{(j)}(n_y)|^2\right)^2,
\end{equation}
and the spectrum-averaged value is
\begin{equation}
\overline{\mathrm{IPR}}=\frac{1}{2q}\sum_{j=1}^{2q}\mathrm{IPR}_{j}.
\end{equation}

\begin{figure*}[t]
\centering
\subfigure{
  \begin{minipage}[b]{0.32\linewidth}\label{fig3a}
    \centering
    \includegraphics[width=1\linewidth]{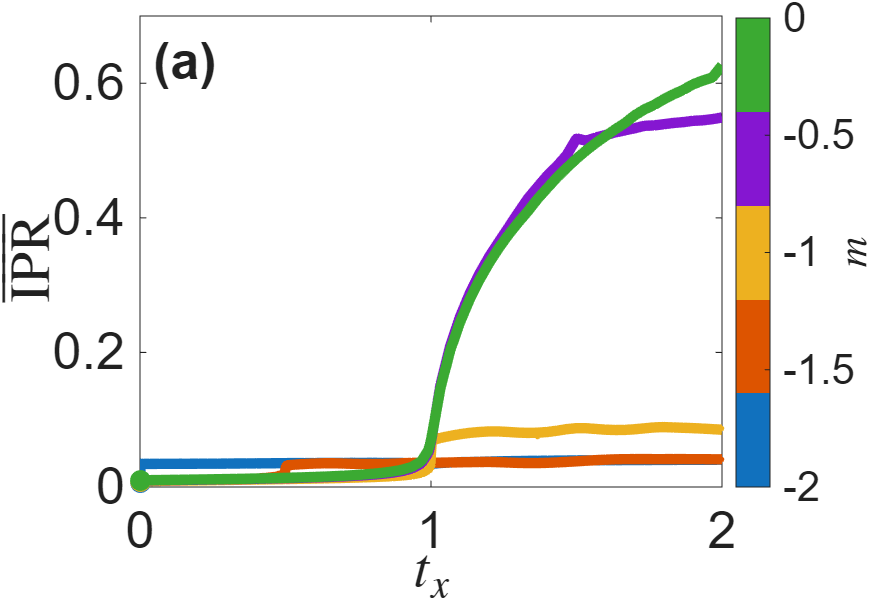}
  \end{minipage}
}
\subfigure{
  \begin{minipage}[b]{0.29\linewidth}\label{fig3b}
    \centering
    \includegraphics[width=1\linewidth]{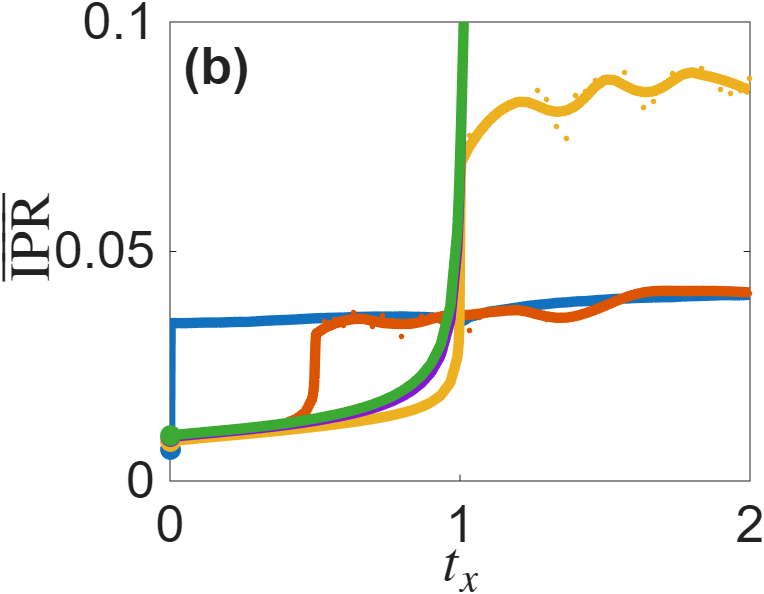}
  \end{minipage}
}
\subfigure{
  \begin{minipage}[b]{0.33\linewidth}\label{fig3c}
    \centering
    \includegraphics[width=1\linewidth]{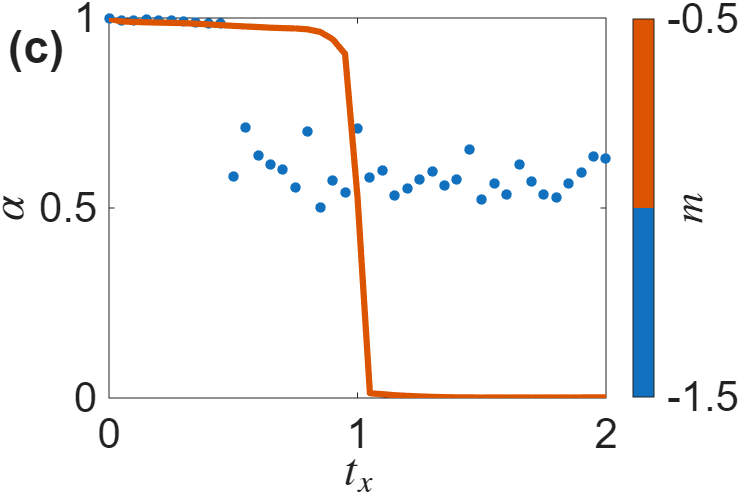}
  \end{minipage}
}
\caption{[(a),(b)] \(\overline{\mathrm{IPR}}\) vs \(t_x\) for different \(m\) with system size \(q = 144\) and \(t_y=1\). (a) For \(m > -1\), \(\overline{\mathrm{IPR}}\) exhibits a sharp increase at \(t_x = 1\), marking the transition from extended (\(t_x < 1\)) to localized (\(t_x > 1\)) states via critical states at \(t_x = 1\). (b) Enlarged view for \(-2 \leqslant m \leqslant -1\), showing that the transition point is given by \(t_x = m + 2\), and the eigenstates remain critical after the transition without localization. 
(c) The scaling index \(\alpha\) vs \(t_x\) for different \(m\). Three distinct scaling behaviors of \(\overline{\mathrm{IPR}}\) are observed for the conventional Anderson transition with \(m=-0.5\), 
whereas only two types are observed for the novel transition with \(m=-1.5\), with the transition point \(t_x=m+2\), featuring a more general anomalous power-law decay.}
\end{figure*}

For \(m > -1\), \(\overline{\mathrm{IPR}}\) exhibits a sharp increase at \(t_x = 1\) [Fig.~\ref{fig3a}]. 
For \(t_x < 1\), the eigenstates have low \(\overline{\mathrm{IPR}}\) and are extended; for \(t_x > 1\), the eigenstates have high \(\overline{\mathrm{IPR}}\) and are localized. 
The transition point \(t_x = 1\) corresponds to critical states.

In Fig.~\ref{fig3b}, we show an enlarged view of Fig.~\ref{fig3a} for \(-2 \leqslant m \leqslant -1\). 
This behavior is distinctly different from the conventional extended-critical-localized phase transitions. 
It illustrates that the transition point of \(\overline{\mathrm{IPR}}\) depends on \(m\), and the value of \(\overline{\mathrm{IPR}}\) after the transition remains much lower than that of the localized phase. 
In fact, for this novel type of eigenstate transition, the transition point is given by \(t_x = m + 2\), 
and the eigenstates remain in the critical phase after the transition without undergoing localization even for a sufficiently strong onsite quasiperiodic potential.

We further investigate the scaling of \(\overline{\mathrm{IPR}}\) with system size \(q\) to obtain a more quantitative characterization of the different eigenstates and their transitions. 
For an extended state, \(\overline{\mathrm{IPR}}\sim q^{-1}\); for an exponentially localized state, \(\overline{\mathrm{IPR}}\) approaches a nonzero constant; and for a critical state, \(\overline{\mathrm{IPR}}\sim q^{-\alpha}\) with \(0<\alpha<1\)~\cite{PhysRevLett.84.3690,PhysRevLett.131.060404,PhysRevLett.72.713,PhysRevLett.131.186303,Aulbach_2004}. We therefore fit the finite-size data to
\begin{equation}\label{IPR_scaling}
\overline{\mathrm{IPR}}=c q^{-\alpha}+\overline{\mathrm{IPR}}_{\infty},
\end{equation}
where the scaling exponent \(\alpha\) is the fractal dimension of the wave function~\cite{PhysRevLett.69.695,PhysRevLett.79.1959}, and \(\overline{\mathrm{IPR}}_{\infty}\) is the average IPR in the thermodynamic limit \(q \to \infty\). 
In the delocalized regime, \(\overline{\mathrm{IPR}}_{\infty} = 0\); in the localized regime, \(\overline{\mathrm{IPR}}_{\infty} > 0\).

We calculate \(\overline{\mathrm{IPR}}\) versus \(q\) for the two types of eigenstate transitions with different \(m\); the fitted scaling exponents \(\alpha\) are presented in Fig.~\ref{fig3c}. 
In all fits, the same Fibonacci sequence of rational approximation and the same sampling of the phase \(k_x\) should be used for each parameter set. 
For the localized states, the finite-size scaling index \(\alpha = 0\), meaning the average IPR is independent of the system size \(q\). In this case, the two constants in Eq.~\eqref{IPR_scaling} can be merged, and the average IPR converges directly to a finite thermodynamic limit value \(\overline{\mathrm{IPR}}_{\infty}\).

For the conventional case \(m=-0.5\), the fitted exponent changes from \(\alpha\approx1\) for \(t_x<1\) to an intermediate value at \(t_x=1\), and finally to \(\alpha\approx0\) for \(t_x>1\). For the persistent-critical case \(m=-1.5\), \(\alpha\) drops from \(\approx1\) at \(t_x=0.5\) but remains between 0 and 1 for all larger \(t_x\) studied. This scaling is the central numerical evidence that the post-transition phase is critical rather than localized.

The scaling exponent \(\alpha\) provides an effective quantitative criterion for distinguishing different eigenstates and characterizing the eigenstate transitions. Interestingly, critical states characterized by the anomalous power-law decay of \(\overline{\mathrm{IPR}}\) appear extensively in the QWZ model under a magnetic field.

\subsection{Physical interpretation}
The two transition patterns can be understood from the competition among the three scales \(|m+2|\), \(t_x\), and \(t_y\). When \(|m+2|>t_y\), the quasiperiodic modulation controlled by \(t_x\) competes primarily with the hopping \(t_y\), producing the usual extended-critical-localized sequence at \(t_x=t_y\). When \(|m+2|\leqslant t_y\), the matrix structure of the onsite modulation changes this competition: the first transition occurs when \(t_x\) reaches \(|m+2|\), but the subsequent phase is protected from exponential localization by the vanishing Lyapunov exponent discussed below. Thus the appropriate interpretation is not merely that the mass term ``suppresses'' localization, but that the matrix quasiperiodic structure and duality constrain the Lyapunov exponent to vanish in the persistent-critical regime.

\section{\label{sec4}Duality of the QWZ model under a magnetic field}
Based on Sec.~\ref{sec3}, we construct the phase diagram of the QWZ model 
under a magnetic field in real space, as shown in Fig.~\ref{fig1}, where 
the extended, localized, and critical phases correspond to regions E, L, and C, respectively. At all phase boundaries, 
the eigenstates are always critical. In this section, we study the duality 
of the QWZ model under a magnetic field. The results reveal a mapping 
between the phase diagrams in real space and its dual space (momentum space)~\cite{aubry1980analyticity,PhysRevLett.131.186303}. 
We then combine the duality with the Lyapunov exponents to explain the 
emergence of different eigenstates.

\subsection{Dual transformation}
We use the dual transformation
\begin{equation}
\psi(n_y)=D e^{ik_y n_y}\sum_{s}\tilde{\psi}(s)e^{is(2\pi\beta n_y+k_x)},
\end{equation}
where \(D=(I-i\sigma_z)/\sqrt{2}\). This unitary rotation keeps the onsite mass term in the same matrix channel after the transformation. Substituting the expansion into Eq.~\eqref{eigeneq} gives the dual equation
\begin{equation}\label{dual_eq}
E\tilde{\psi}(s)=\tilde{U}(s)\tilde{\psi}(s)-\tilde{T}\tilde{\psi}(s+1)-\tilde{T}^{\dagger}\tilde{\psi}(s-1),
\end{equation}
with
\begin{widetext}
\begin{equation}
\tilde{U}(s)=\begin{pmatrix}
m+2-t_y\cos(k_y+2\pi\beta s) & t_y\sin(k_y+2\pi\beta s) \\
t_y\sin(k_y+2\pi\beta s) & -[m+2-t_y\cos(k_y+2\pi\beta s)]
\end{pmatrix},
\end{equation}
\end{widetext}
and \(\tilde{T}=\dfrac{t_x}{2}\begin{pmatrix}1&1\\-1&-1\end{pmatrix}\).
The dual model has the same structure as the real-space model under the exchange
\begin{equation}
(t_x,k_x)\longleftrightarrow(t_y,k_y).
\end{equation}
Thus the line \(t_x=t_y\) is self-dual up to the exchange of Bloch phases. More generally, the transformation maps the real-space phase diagram to the momentum-space phase diagram: extended and localized regions are exchanged, whereas critical regions remain delocalized in both representations.

Since \(t_x\) and \(t_y\) are exchanged under the dual transformation, we can establish a mapping between the eigenstates in real space and momentum space. 
In the momentum-space phase diagram, the horizontal axis is \(t_y/t_x\) and the vertical axis is \(|m+2|/t_x\). The extended phase in 
real space is mapped to the localized phase in momentum space, while the localized phase in real space is mapped to the extended phase in momentum space. 
More specifically, we further classify the two types of eigenstate transitions. Interestingly, the extended phase in real space with \(|m+2| > t_y\) is mapped to a subregion of the localized phase in momentum space, rather than the entire localized region. Meanwhile, the extended phase in real space with \(|m+2| \leqslant t_y\) is mapped to the remaining region of the localized phase in momentum space.

In contrast, the critical phase exhibits completely different behavior. 
It is delocalized in both real and momentum spaces. Under the dual 
transformation, the critical phase in real space with \(|m+2|<t_x<t_y\) is mapped to the 
critical phase on the right side of the momentum-space phase diagram, while the critical phase in 
real space with \(|m+2|<t_y<t_x\) is mapped to the critical phase on the left side of the momentum-space phase diagram. 
From a graphical perspective, the entire critical phase region can be regarded as an extension 
of the critical phase at \(t_x=t_y\), and the critical phases on both sides are dual to each other with respect to \(t_x=t_y\).

\subsection{Lyapunov-exponent characterization}
The duality becomes predictive once it is combined with a Lyapunov-exponent calculation. A more detailed discussion is given in Appendix~\ref{app:lyapunov}. We define the Lyapunov exponent as a nonnegative inverse localization length~\cite{D_J_Thouless_1972}. In the localized region \(|m+2|>t_y\) and \(t_x>t_y\), the transfer matrix gives a positive real-space Lyapunov exponent, and the eigenstates are exponentially localized. The dual region is therefore extended in real space and localized in momentum space. At the separatrix \(t_x=t_y<|m+2|\), the Lyapunov exponent vanishes and the states are critical. In the regime \(|m+2|\leqslant t_y\), the Lyapunov exponent vanishes throughout the post-transition region \(t_x\geqslant |m+2|\). Because the dual representation also has a vanishing Lyapunov exponent, the states are delocalized in both real and momentum spaces, which identifies them as critical rather than extended or localized.

\section{\label{sec5}Dynamics of wave packets and transport properties}
In this section, we study the time evolution of wave packets along the \(y\)-direction in the QWZ model under a magnetic field. 
Owing to the more extensive presence of the critical phase, richer anomalous power-law quantum diffusion is expected to emerge in the QWZ model.

\begin{figure*}[t]
\centering
\subfigure{
  \begin{minipage}[b]{0.32\linewidth}\label{fig4a}
    \centering
    \includegraphics[width=1\linewidth]{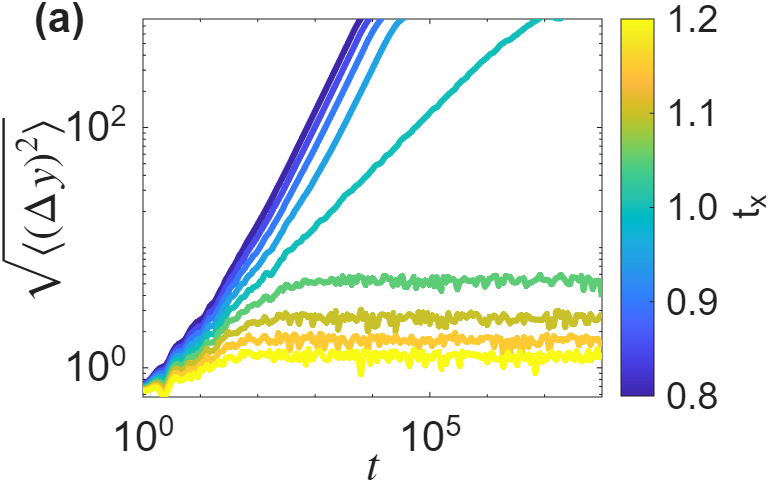}
  \end{minipage}
}
\subfigure{
  \begin{minipage}[b]{0.32\linewidth}\label{fig4b}
    \centering
    \includegraphics[width=1\linewidth]{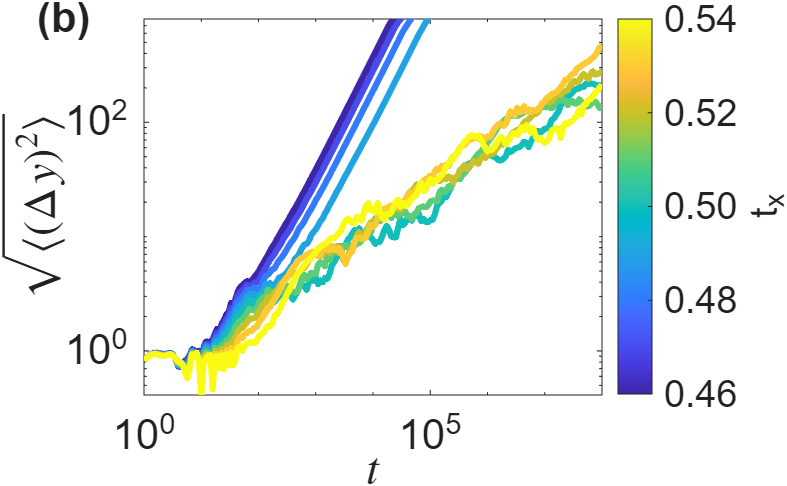}
  \end{minipage}
}
\subfigure{
  \begin{minipage}[b]{0.3\linewidth}\label{fig4c}
    \centering
    \includegraphics[width=1\linewidth]{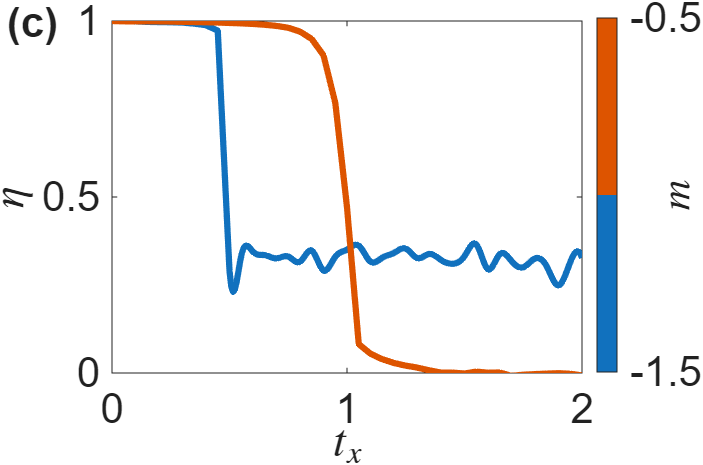}
  \end{minipage}
}
\caption{Time evolution of wave packets and power-law indices for the QWZ model. We set \(q = 4181\) and \(t_y=1\). 
[(a),(b)] Wave packet width \(\sqrt{\langle (\Delta y)^2 \rangle}\) vs time \(t\) for (a) \(m=-0.5\) and (b) \(m=-1.5\) with different \(t_x\). 
(c) Power-law index \(\eta\) vs \(t_x\) for different \(m\). 
For \(m=-0.5\), \(\eta \approx 1\) (ballistic transport) for \(t_x < 1\), drops to 0.47 (anomalous diffusion) at \(t_x=1\), and then to \(0\) (localized) for \(t_x > 1\). 
For \(m=-1.5\), \(\eta \approx 1\) (ballistic transport) for \(t_x < m+2\), drops to \(\approx 0.3\) at \(t_x=m+2\), and remains around 0.3 (anomalous diffusion) for \(t_x > m+2\).}
\end{figure*}

Suppose an electron is initially localized at a lattice site \(n_y = n_0\) in one of the atomic orbitals (\(A\) or \(B\)). The time evolution of the wave function follows the time-dependent Schrödinger equation (with \(\hbar = 1\))
\begin{eqnarray}
i\dfrac{\partial}{\partial t}\Psi(n_y,t)&=&U(n_y) \Psi(n_y,t)- T_y \Psi(n_y+1,t)\nonumber\\
&&- T_y^\dagger \Psi(n_y-1,t).
\end{eqnarray}
where \(\Psi(n_y,t) = \begin{pmatrix} \Psi_A(n_y,t) \\ \Psi_B(n_y,t) \end{pmatrix}\) is a superposition of eigenstates multiplied by their dynamical phase factors:
\begin{equation}
\Psi(n_y,t)=\sum_{j=1}^{2q}d_j\psi^{(j)}(n_y)e^{-iE_j t},
\end{equation}
where the coefficients \(d_j=\sum_{n_y}[\psi^{(j)}(n_y)]^\dagger\Psi(n_y,0)\), obtained by projecting the initial condition onto the eigenstates. The index \(j\) on \(E_j\) is essential because each eigenstate accumulates its own dynamical phase.

The root-mean-square width is evaluated as
\begin{equation}
\sqrt{\langle(\Delta y)^2\rangle}=\sqrt{\sum_{n_y}d_q(n_y,n_0)^2|\Psi(n_y,t)|^2},
\end{equation}
where \(|\Psi(n_y,t)|^2 = |\Psi_A(n_y,t)|^2 + |\Psi_B(n_y,t)|^2\) and \(d_q(n_y,n_0)\) is the shortest distance on a ring of length \(q\) under periodic boundary conditions. The fitting time window must be restricted to times before the wave packet reaches the boundary, and \(d_q(n_y,n_0)\) is replaced by \(|n_y-n_0|\) for open boundaries. We fit \(\sqrt{\langle(\Delta y)^2\rangle}\sim t^\eta\)~\cite{HiramotoKohmoto1992,PhysRevA.36.5349,1988230,doi:10.1143/JPSJ.57.1365,PhysRevLett.79.1959,PhysRevE.86.021136} over an intermediate asymptotic window before finite-size recurrences, where \(\eta\) is the power-law index associated with transport properties, 
which can provide a quantitative description of electron dynamics before and after the two types of eigenstate transitions.

For \(m = -0.5\), the time evolution of the wave packets for different \(t_x\) is shown in Figs.~\ref{fig4a} and~\ref{fig4c}, 
which is consistent with the conventional Anderson transition in the AAH model~\cite{doi:10.1143/JPSJ.57.1365,HiramotoKohmoto1992}. 
For the extended states with \(t_x < 1\), \(\sqrt{\langle (\Delta y)^2 \rangle}\) increases rapidly and tends to infinity for an unbounded system with the power-law index \(\eta = 1\), corresponding to ballistic transport. 
In contrast, for the localized states with \(t_x > 1\), \(\sqrt{\langle (\Delta y)^2 \rangle}\) remains finite and small, and the wave packet does not 
spread even as \(t \to \infty\), with the power-law index \(\eta = 0\). 
Intermediately, for the critical states at \(t_x = 1\), \(\sqrt{\langle (\Delta y)^2 \rangle}\) grows more slowly than that for the extended states, exhibiting anomalous power-law quantum diffusion with the power-law index \(\eta = 0.47\), a hallmark of subdiffusion.

For \(m=-1.5\), the time evolution of the wave packets for different \(t_x\) is shown in Figs.~\ref{fig4b} and~\ref{fig4c}, illustrating the dynamical behaviors of the wave packets before and after the novel eigenstate transition. For the extended states with \(t_x < m+2\), \(\sqrt{\langle (\Delta y)^2 \rangle}\) increases rapidly with the power-law index \(\eta = 1\), indicating ballistic motion. Beyond the transition point, i.e., \(t_x \geqslant m+2\), the wave function remains in the critical phase as \(t_x\) increases, and the wave packet width \(\sqrt{\langle (\Delta y)^2 \rangle}\) grows with an anomalous power-law index \(\eta \approx 0.3\), 
greatly enriching the possibilities for realizing anomalous diffusion. Although the onsite quasiperiodic potential is strong enough, it does not lead to localization in this case. However, the power-law index \(\eta\) is generally smaller than that of the critical states in the conventional Anderson transition, indicating that the tendency toward localization induced by the quasiperiodic potential still exists but is suppressed by the condition \(|m+2| \leqslant 1\).

The wave-packet dynamics are consistent with the eigenstate classification. Extended states show ballistic spreading with \(\eta\approx1\), localized states show saturation with \(\eta\approx0\), and critical states show anomalous power-law spreading with \(0<\eta<1\). In the persistent-critical regime, \(\eta\) remains finite but subdiffusive for \(t_x\geqslant|m+2|\), confirming that strong quasiperiodic modulation does not produce ordinary localization in this parameter range.

\section{Summary and outlook}\label{sec6}
We have studied eigenstate transitions, duality, and wave-packet dynamics in the Qi-Wu-Zhang Chern-insulator model under an irrational magnetic flux. The model reduces to a spinor quasiperiodic chain whose matrix onsite modulation is controlled by \(t_x\). When \(|m+2|>t_y\), the model exhibits the conventional extended-critical-localized sequence with a critical line at \(t_x=t_y\). When \(|m+2|\leqslant t_y\), it instead exhibits a persistent critical phase: the states change from extended to critical at \(t_x=|m+2|\) and remain critical for stronger modulation.

The average inverse participation ratio provides the main static diagnostic. Its finite-size scaling gives \(\alpha\approx1\) in extended regions, \(\alpha\approx0\) in localized regions, and \(0<\alpha<1\) in critical regions. A dual transformation exchanging \(t_x\) and \(t_y\), together with a Lyapunov-exponent analysis, explains why the post-transition region for \(|m+2|\leqslant t_y\) is critical rather than localized. Wave-packet dynamics provide an independent dynamical signature: extended, critical, and localized states correspond respectively to ballistic motion, anomalous diffusion, and saturation.

A direct topological characterization remains an important next step. In particular, Chern-number or Bott-index calculations across the eigenstate transition would clarify whether the persistent critical phase is tied to band exchange, anomalous open-orbital subbands, or a distinct localization mechanism of the magnetic QWZ model.

\begin{acknowledgments}
This work was financially supported by the National Key Research and Development Program of China (Grant No. 2024YFA1409001), the National Natural Science Foundation of China (Grants No. 12374037), and the Fundamental Research Funds for the Central Universities.
\end{acknowledgments}

\appendix

\section{Analytical limit without the influence of the magnetic field}\label{app:no_field}

In the special limit \(t_x=0\) and \(m=-2\), the onsite potential \(U(n_y)\) in Eq.~\eqref{eigeneq} vanishes identically. The eigen-equation can then be solved exactly, yielding the two eigenvalues \(E=\pm t_y\), with corresponding eigenstates
\begin{equation}\label{eqA1}
\psi_+= \frac{e^{ik_y n_y}}{\sqrt{q}}\begin{pmatrix} -i \sin \dfrac{k_y}{2}  \\ \cos \dfrac{k_y}{2}  \end{pmatrix},
\psi_-= \frac{e^{ik_y n_y}}{\sqrt{q}}\begin{pmatrix} \cos \dfrac{k_y}{2}  \\ -i \sin \dfrac{k_y}{2}  \end{pmatrix}.
\end{equation}
In this limit, the system is physically equivalent to the case without the influence of the magnetic field: the effective onsite potential vanishes, and the eigenstates become extended Bloch waves along the \(y\)-direction. 
Substituting Eq.~\eqref{eqA1} into the IPR definition given in Eq.~\eqref{IPR} yields 
\(\overline{\mathrm{IPR}} = \dfrac{1}{2} (\mathrm{IPR}_+ + \mathrm{IPR}_-) = \dfrac{1}{q}\), confirming the extended nature of the eigenstates.

In this limit, the Hamiltonian becomes translationally invariant along the \(y\)-direction, and the eigenstates \(\psi_\pm\) are Bloch waves with dispersion \(E_\pm=\pm t_y\). An initially localized wave packet therefore spreads ballistically, with the root-mean-square width growing linearly in time, \(\sqrt{\langle(\Delta y)^2\rangle}\approx t\), corresponding to the power-law index \(\eta=1\).

On a finite ring of length \(q\), this ballistic spreading is eventually interrupted by the wave packet wrapping around the ring, leading to periodic revivals and an upper bound of the width. The ballistic transport is nevertheless recovered in the thermodynamic limit \(q\to\infty\) or at times before the wave packet reaches the boundary.

This behavior is the standard non-magnetic benchmark for the QWZ model. It provides a clear contrast with the anomalous diffusion observed in the critical phases discussed in the main text, confirming that the nontrivial power-law dynamics in the QWZ model are genuinely induced by the magnetic quasiperiodic potential.

\section{Lyapunov-exponent analysis}\label{app:lyapunov}
Without loss of generality, we set \(t_y = 1\). Combining the Thouless formula with the relationship between the densities of states in real and momentum spaces, for an irrational \(\beta\), the Lyapunov exponent satisfies the dual relation~\cite{aubry1980analyticity}
\begin{equation}
\gamma = \tilde{\gamma} + \ln \lambda,
\end{equation}
where \(\gamma\) and \(\tilde{\gamma}\) are the Lyapunov exponents in real and momentum spaces, respectively, which are independent of energy for the QWZ model due to the absence of a mobility edge. 
According to Avila's global theory, the correlation factor \(\lambda\) takes the following form~\cite{10.1007/s11511-015-0128-7,ZHOU20261654}
\begin{equation}
\lambda = 
\begin{cases} 
1, &t_x \leqslant 1 \\ 
1, &|m+2| \leqslant 1 <t_x \\ 
|m+2| + \sqrt{(m+2)^2 - 1}, &1 < |m+2| < t_x \\ 
t_x \dfrac{|m+2| + \sqrt{(m+2)^2 - 1}}{|m+2| + \sqrt{(m+2)^2 - t_x^2}}, &1<t_x \leqslant |m+2|
\end{cases}
\end{equation}  

When both \(|m+2|>1\) and \(t_x>1\), we have \(\lambda > 1\) and \(\tilde{\gamma} \geqslant 0\), thus \(\gamma > 0\), 
implying that all eigenstates are exponentially localized. This corresponds 
to region L of the phase diagram in Fig.~\ref{fig1} with the localization length \(\xi = \gamma^{-1} = (\ln \lambda)^{-1}\). 
Via the dual transformation, we obtain that region E corresponds to the extended phase with the correlation length \(\tilde{\xi} = \tilde{\gamma}^{-1} = -(\ln \lambda)^{-1}\), 
obtained from \(\gamma =\tilde{\gamma} + \ln \lambda = 0\). 
At the separatrix between the extended and localized phases, i.e., \(t_x =1< |m+2|\), we have \(\gamma = \tilde{\gamma} = 0\), and the eigenstates are critical.

For \(|m+2| \leqslant 1\) and \(t_x \geqslant |m+2|\), 
we always have \(\lambda=1\) and \(\gamma = \tilde{\gamma} = 0\), 
which implies that the wave functions are delocalized in both real and momentum spaces, corresponding to the critical phase. The dual transformation also reveals that the critical phases on both sides are dual to each other with respect to \(t_x = 1\), indicating the robustness of the critical phase.

Based on the above discussion, the phase diagram of the QWZ model under a magnetic field is fully determined, covering all possible eigenstates and their transitions.

\bibliography{TengmingLou_ref}

\end{document}